# SECURE IMAGE STEGANOGRAPHY USING CRYPTOGRAPHY AND IMAGE TRANSPOSITION

Khan Muhammad[1], Jamil Ahmad[2], Muhammad Sajjad[3], Muhammad Zubair[4]

## ABSTRACT

Information security is one of the most challenging problems in today's technological world. In order to secure the transmission of secret data over the public network (Internet), various schemes have been presented over the last decade. Steganography combined with cryptography, can be one of the best choices for solving this problem. This paper proposes a new steganographic method based on gray-level modification for true colour images using image transposition, secret key and cryptography. Both the secret key and secret information are initially encrypted using multiple encryption algorithms (bitxor operation, bits shuffling, and stego key-based encryption); these are, subsequently, hidden in the host image pixels. In addition, the input image is transposed before data hiding. Image transposition, bits shuffling, bitxoring, stego key-based encryption, and gray-level modification introduce five different security levels to the proposed scheme, making the data recovery extremely difficult for attackers. The proposed technique is evaluated by objective analysis using various image quality assessment metrics, producing promising results in terms of imperceptibility and security. Moreover, the high quality stego images and its minimal histogram changeability, also validate the effectiveness of the proposed approach.

[1] PhD student, Digital Contents Research Institute, Sejong University, Seoul, Korea, Ph. +82-010-48312104, Fax: +82-02-3408-4339, Email: khanmuhammad@sju.ac.kr.
[2] PhD student, Digital Contents Research Institute, Sejong University, Seoul, Korea, Ph. +82-010-47862016, Fax: +82-02-3408-4339, Email: jamilahmad@sju.ac.kr.
[3] Research Associate, Department of Computer Science, Islamia College Peshawar, Pakistan, Ph. +92-333-9319519, Fax: +82-02-3408-4339, Email: muhammad.sajjad@icp.edu.pk.
[4] Lecturer, Department of Computer Science, Islamia College Peshawar, Pakistan, Ph. +92-333-9131479, Fax: +82-02-3408-4339, Email: zubair@icp.edu.pk.

| | |
|---|---|
| 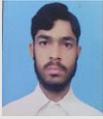 | *Khan Muhammad received his Bachelors in computer science from Islamia College, Peshawar, Pakistan in 2014. He is studying PhD in digital contents at Sejong University, Seoul, South Korea. His research interests include image processing, data hiding, steganography, watermarking and video summarization.* |
| 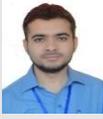 | *Jamil Ahmad received his Bachelors and Master in computer science, respectively, from the University of Peshawar, Pakistan and Islamia College, Peshawar, Pakistan. He is studying PhD in digital contents at Sejong University, Seoul, Korea. His research interests include image analysis, semantic image representation and content based multimedia retrieval.* |
| 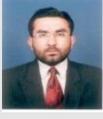 | *Muhammad Sajjad is a Research Associate at Islamia College, Peshawar, Pakistan. He received his PhD in digital contents from Sejong University, Seoul, South Korea. His research interests include digital image super-resolution and reconstruction, sparse coding, video summarisation and prioritisation, image/video quality assessment, and image/video retrieval.* |
| 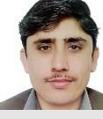 | *Muhammad Zubair is a Lecturer in the Department of Computer Science at Islamia College, Peshawar, Pakistan. He received his Bachelors in computer science from University of Peshawar, Pakistan in 2009. His research interests include network security, image processing, quality of service and IP based routing.* |

# 1. INTRODUCTION

Steganography is the process of writing covert messages so that its existence cannot be detected using human visual system (HVS) [1, 2]. The most important prerequisites of steganography include an input image, secret information and data hiding algorithm. To increase the security up to some extent, sometimes a stego key and encryption procedure is also used along with steganographic algorithm. Steganography can be used for a number of different applications including secure exchange of top-secret messages between sensitive organisations, securing online banking, and voting systems [3-6]. It can be one of the most nefarious ways for attackers to send viruses and Trojan horses. Furthermore, terrorists and criminals can use it for secret communication. A number of different steganographic techniques based on carrier object exists including text based methods, image steganographic methods, video and audio based data hiding and network packets based data hiding schemes. [1, 7, 8].

Steganographic techniques are divided into the following two categories.

1) Spatial domain techniques which are direct modification of input image's pixels including least significant bit (LSB) methods [9-12], edges based methods[13-17], pixel-value-differencing (PVD) based methods[18, 19], and pixel-indicator-techniques (PIT) [20-27]. These methods result in high quality stego images and provide higher payload but are vulnerable to different normal attacks such as joint photographic experts group (JPEG) compression, noise attacks, and low-pass/high-pass filtering [28] and geometric attacks such as image resizing, cropping and rotations by different angles [29, 30].

2) Transform domain techniques use the transformed-coefficients of the input image obtained through different transforms such as discrete Fourier transform (DFT) [31], discrete wavelet transform (DWT) [32], discrete cosine transform (DCT) [33, 34] and contourlet transform [35] for data hiding. These methods have lower payload but can survive against different attacks.

In this area of steganography, two different groups are working. The first group designs steganographic algorithms while the second group develops its counter attacks i.e. Steganalysis. Steganalysis is the science of defeating steganography in a battle that will not end at all. Steganalysis can be active when the embedded data is to be retrieved. Alternatively, it can be passive where the interest is in detecting the secret information. Detecting the steganography is an important issue for law enforcement authorities as criminals and terrorists mostly use steganography for information interchange [4, 5, 36, 37].

In this paper, a new colour image steganographic technique has been proposed for information hiding. Colour image has been used as a carrier object because it contains more redundant bits. The main contributions of this research work are as follows

1) A new image steganographic technique using gray-level modification and cryptography.

2) Better quality of stego images as compared to other state-of-the-art techniques, verified by experimental results, reducing the chances of detection by adversaries.

3) Encryption of secret key and secret data before data hiding, increasing the security.

The rest of the paper is organized as follows. In section 2, some well-known steganographic techniques are briefly described that are related to the proposed work. Section 3 explains the proposed work in detail, followed by experimental results and discussion in section 4. The conclusion of the paper and future suggestions are presented in section 5.

## 2. LITERATURE REVIEW

The usage of steganography was started by Greeks with the famous story of shaved head. Over the last few decades, different techniques have been used for message hiding such as tablets with wax, microdots, invisible ink, semagrams, and open codes. In digital steganography, the basic technique of data hiding is to replace the LSBs of the input image with the bits of secret data as described in [38] and its basic idea is given as under

Binary representation of eight (8) pixels: 10001101, 10000010, 01110110, 01100001, 00101000, 10000100, 01001011, 01110111.

Secret character: **A**     01000001

After hiding this secret character (**A**) in these pixels, the pixel values in binary format are obtained as follows: 1000110**0**, 1000001**1**, 01110110, 0110000**0**, 00101000, 10000100, 0100101**0**, and 01110111.

The bold face LSBs indicate the changed pixels during data hiding. It can be seen in the above that only four pixels change which shows approximately half of the pixels change. Therefore, the distortion caused by this approach in stego images is almost undetectable using HVS.

LSB matching (LSB-M) [17] is a modified version of LSB method which adds/subtracts unity to/from the pixel value if its LSB is not identical to a given secret bit. The asymmetric artifacts produced by LSB method and LSB-M method are reduced by LSB-M revisited (LSB-MR) [12]. Also LSB-MR method interprets the pixel values dependently by considering the relationship between nearby pixels and minimizes the modification rate from 0.5 to 0.325 in the unit bits per pixel (bpp). The extraction of secret data embedded through LSB, LSB-M and LSB-MR is relatively easy for an attacker which is one of the shortcomings of these methods.

To make the extraction of data difficult for attackers, the authors in [39] proposed stego colour cycle (SCC) method that scatters secret data in three channels of the input cover image in a cyclic order. The data is embedded in the sequence of red, green, blue and so on. SCC is further improved by authors in [8] using randomisation. The aforementioned algorithms are better than LSB, LSB-M and LSB-MR methods as it scatters the secret data in different channels of the input cover image. Nevertheless, successfully extracting data from a few pixels can compromise these methods.

Karim et al. [40] proposed a new approach to enhance the robustness of existing LSB substitution method by adding one level security of secret key. In the proposed method, secret key and red channel are used as an indicator while green and blue channels are used as data channels. On the basis of secret key bits and red channel LSBs, the secret data bits are embedded either in green channel or blue channel. An intruder cannot easily extract the secret information without the correct secret key. Moreover, the experimental results also show better image quality and robustness.

The methods discussed so far produce stego images of low quality which are easily detectable using HVS. Furthermore, the data is embedded in cover images without encryption which makes its extraction easy for attackers. This presented research work solves these problems and proposes a new scheme which improves the quality of stego images in addition to increasing the security of secret data during transmission.

## 3. THE PROPOSED METHODOLOGY

The proposed method is a new robust approach to map secret data to one of the three channels of the RGB image. The proposed method uses the idea of transposition, bitxoring, bits shuffling, secret key, and cryptography to design an advanced steganographic system. Unlike other methods, the proposed method have the following multiple security levels.

1) All the three channels of the input carrier image are transposed before they can be used to map secret data in order to deceive the attacker.
2) The secret key and secret data is encrypted using multiple encryption algorithms that are applied on it one after another.
3) Secret data is mapped to blue channel of the carrier image using gray-level modification method (GLM).

The proposed method uses two different modules named as encryption module and mapping module in order to hide secret data to the carrier image pixels. The overall diagrammatic representation of the proposed framework is shown in **Figure 1**. The modules of the proposed algorithm are briefly discussed in the forthcoming sections.

### 3.1 Encryption Module

This module is responsible to encrypt both the secret key and secret data. The final output of this module is secret key and secret data bits in encrypted form. This module performs the following operations on secret key and secret data.

1) Select the secret data and a suitable secret key for encryption
2) Convert the secret key into one-dimensional (1-D) array of bits
3) Apply the bitxor operation on these bits with logical 1.
4) Shuffle these encrypted bits such that the bits with even and odd indices are interchanged.
5) *If secret key bit =1*

*Then perform bitxor operation of secret message bit with logical 1.*

*Else*

*Do not perform bitxor operation.*

*End if*

6) Repeat step 4 until all secret data bits are encrypted.

## 3.2 Mapping Module

This module is responsible for mapping the secret encrypted data into the carrier image pixels. Before mapping, the carrier image channels are transformed and then a 1-1 mapping between secret data bits and image pixels is maintained. The end result of this module is a stego image, containing secret information.

## 3.3 Embedding Algorithm

*Input:* Cover colour image, secret key, and secret data

*Output:* Stego image

1) Select the colour cover image and divide it into red, green, and blue channels
2) Apply image transpose on all the three channels of the input image
3) Encrypt the secret key and secret data according to the encryption module 3.1
4) *If the first bit of secret data=1*

   *Then convert all pixel values of blue channel to odd by adding 1*

   *Else*

   *Convert all pixel values of blue channel to even by adding 1*

5) Map the secret data of step 4 based on secret key bits (SKB) such that

   *If SKB=0 && pixel value=even OR SKB=1 && pixel value=odd*

   `   *Then leave the pixel unchanged*

   *Else if SKB=0 && pixel value=odd*

   *Then subtract 1 from pixel value*

> *Else if SKB=1 && pixel value=even*
>
> *Then add 1 to pixel value*

6. Repeat step 5 until all secret bits are mapped with the gray-levels of carrier image

7. Take the transpose of all three planes and combine them to make the stego image

**3.4 Extraction Algorithm**

*Input:* Stego image, secret key

*Output:* Secret data

1) Select the colour stego image and divide it into red, green, and blue channels
2) Apply image transpose on all the three channels of the stego image
3) Extract LSB of the blue channel
4) Repeat step 3 until all secret bits are successfully extracted
5) Decrypt these bits by applying the reverse method of encryption module 3.1 to get the original text

**4. EXPERIMENTAL RESULTS AND ANALYSIS**

This section presents the experimental results based on various image quality assessment metrics for performance evaluation. The proposed method is compared with the Karim et al. method [40] and are implemented using MATLAB R2013a. The evaluation is done using multiple experiments from different perspectives on different standard colour images of varying dimensions. For example, one experiment is to embed a text file of eight kilobyte (8KB) in different standard colour images of dimension (256×256) like Lena, baboon, peppers, army, airplane, building, and house. Another experiment is to embed different amount of data in one standard image of the same dimension. The third experiment is to embed the same amount of data in the same image but with different dimensions. For comparison of the proposed method with the existing methods, both the subjective and objective measurements have been used. HVS is a subjective measurement for identification of obvious distortion in the stego images by naked eye [41]. **Figure 2-5** show some sample standard colour cover images, stego images and their histograms. Using HVS, it can be noted in **Figure 2-5** that the cover and stego images and their histograms are indistinguishable from one another.

Objective analysis has also been used for comparison of proposed method with other methods by calculating the normalized-cross-correlation (NCC), peak-signal-to-noise ratio (PSNR), mean square error (MSE), and root MSE (RMSE). Moreover, to show the better performance of the proposed method, histograms changeability and comparison graphs are also mentioned. PSNR, MSE, RMSE and NCC were calculated by the formulae as given in Eqs. (1)-(4) [2, 42].

$$PSNR = 10\log_{10}\left(\frac{C_{max}^2}{MSE}\right) \quad (1)$$

$$MSE = \frac{1}{MN}\sum_{x=1}^{M}\sum_{y=1}^{N}(S_{xy} - C_{xy})^2 \quad (2)$$

$$RMSE = \sqrt{MSE} \quad (3)$$

$$NCC = \frac{\sum_{x=1}^{M}\sum_{y=1}^{N}(S(x,y) \times C(x,y))}{\sum_{x=1}^{M}\sum_{y=1}^{N}(S(x,y))^2} \quad (4)$$

where $x$ and $y$ are the loop counters; $M$ and $N$ are image dimensions; $C_{max}$ is the maximum value among all pixels of both cover and stego images; $S$ is the stego image; and $C$ shows the cover image [7, 43, 44].

The experimental results of the proposed algorithm and the Karim et al. [40] algorithm are given in **Tables 1-4**. **Table 1** contains the PSNR, MSE, and RMSE scores for both methods. The stego images having PSNR value greater than 40 decibel (dB) are considered to be of high quality. However, PSNR score smaller than 30dB represents lower quality of stego images and hence causes noticeable deformation in stego images which is easily detectable by HVS. The PSNR values for the proposed algorithm are greater than the Karim' et al. [40] algorithm which shows high quality of stego images. Similarly, the MSE values of the proposed algorithm are small as compared to the Karim et al [40] method. Furthermore, the RMSE scores of proposed method are smaller than the Karim et al. [40] method. This means that the proposed algorithm provides promising results in terms of PSNR, confirming its better performance.

The comparison graph of the proposed method and the Karim et al. [40] method is shown in **Figure 6**. The graph is drawn on the basis of fifteen different smooth and edgy images. The PSNR values are shown on the y-axis and image names on the x-axis. The graph clearly shows that there is up and down in the values of PSNR of the Karim et al. [40] method but the values of PSNR in the proposed method are almost the same and do not vary significantly. This verifies that the proposed method performs well for all types of images (edgy and smooth) as compared to the Karim et al [40] method.

**Table 2** shows the comparison of both methods using PSNR with variable amount of cipher that is embedded in the standard colour image (baboon) of the same dimension (256×256). **Table 2** clearly shows that the proposed method gives more PSNR score as compared to the Karim et al. [40] method. Similarly, the comparative analysis graph of both the methods with variable amount of cipher embedded in a standard colour image of the same dimension is shown in **Figure 7**. The graph is drawn on the basis of PSNR values of **Table 2**. The comparative graph of the proposed algorithm as compared to the Karim et al. [40] algorithm clearly shows its better results in terms of PSNR which validate the effectiveness of the proposed method.

**Table 3** provides the comparison of both methods using PSNR with the same amount of cipher embedded and same standard image (baboon) but with different dimensions. The results of **Table 3** show that there is variation in the PSNR score of the Karim et al. [40]

method but the PSNR score of the proposed algorithm is increasing as the image size is increased. Similarly, the comparative graph of both methods using PSNR with variable dimensions, same image and same amount of cipher embedded is also shown in **Figure 8** which vividly describes the effectiveness of the proposed technique.

The similarity between two images can be measured by using the correlation function. NCC is a statistical error metric that has been used to measure the similarity between two digital images in this research work. **Table 4** shows NCC for both the algorithms. If the NCC value is unity, both images become identical to each other. The value of NCC in **Table 4** close to unity shows that both the images are similar and differences are small. **Table 4** clearly shows that the NCC values for the proposed algorithm in all cases are greater than the Karim et al. [40] algorithm. This shows that the proposed algorithm provide better results in terms of NCC also and verifies its effectiveness.

## 5. CONCLUSIONS

In this paper, a new method is proposed to map secret data to the gray-levels of the carrier image by utilising the concepts of transposition, bitxoring, bits shuffling, secret key, and cryptography with high imperceptibility and security. An average PSNR of 58dB, RMSE with 0.6673, and NCC with 0.9917 is achieved using the proposed method which are better than the existing method in the literature with PSNR=40, RMSE=0.8115, and NCC=0.981. The proposed method improved the security as well as the quality of stego images and provided promising results in terms of high PSNR, NCC, and less histogram changeability as compared to existing methods. The distinguishing properties of the proposed algorithm include transposition, bitxoring, and bits shuffling, adding multiple security levels to the proposed method. These different security levels create multiple barriers in the way of an attacker. Therefore, it is difficult for a malicious user to extract the actual secret data.

**ACKNOWLEDGMENT**

The authors are thankful to the anonymous reviewers, associate editor, and editorial board members for providing their useful and constructive comments which improved the quality of this paper. The authors also acknowledge Dr. Zahoor Jan for his continuous help and support.

**REFERENCES**

[1] Chang Y-T, Wu M-H, Wang S-J. Steganalysis to Data Hiding of VQ Watermarking Upon Grouping Strategy. In: Information and Communication Technology, Springer, 2014. p. 633-642.

[2] Sajjad M, Mehmood I, Baik SW. Image Super-resolution Using Sparse Coding Over Redundant Dictionary Based on effective Image Representations. J Visual Commun Image Rep 2015;26(1):50-65.

[3] Qin C, Chang C-C, Chiu Y-P. A Novel Joint Data-Hiding and Compression Scheme Based on SMVQ and Image Inpainting. IEEE Trans Image Process 2014;23(3):969-978.

[4] Cheddad A, Condell J, Curran K, Kevitt PMc. Digital Image Steganography: Survey and Analysis of Current Methods. Sig Process 2010;90(3):727-752.

[5] Hamid N, Yahya A, Ahmad RB, Al-Qershi OM. Image Steganography Techniques: An Overview. Int J Comp Sci Secu 2012;6(3):168-187.

[6] Liao X, Shu C. Reversible Data Hiding in Encrypted Images Based on Absolute Mean Difference of Multiple Neighboring Pixels. J Visual Commun Image Rep 2015;28(3):21-27.

[7] Sajjad M, Ejaz N, Baik SW. Multi-kernel Based Adaptive Interpolation For Image Super-resolution. Multi Tool App 2012;72(3):2063-2085.

[8] Jamil Ahmad NUR, Jan Z, Muhammad K. A Secure Cyclic Steganographic Technique for Color Images using Randomization. Tech J Uni Engg Tech Taxila 2014;19(3):57-64.

[9] Qazanfari K, Safabakhsh R. A New Steganography Method which Preserves Histogram: Generalization of LSB$^{++}$. Info Sci 2014;277(7):90-101.

[10] Yang C-H, Weng C-Y, Wang S-J, Sun H-M. Adaptive Data Hiding in Edge Areas of images With Spatial LSB Domain Systems., IEEE Trans Info Forens Sec 2008;3(3):488-497.

[11] Zhang W, Zhang X, Wang S. A Double Layered "Plus-Minus One" Data Embedding Scheme. Sig Process Let, IEEE 2007;14(11):848-851.

[12] Mielikainen J. LSB Matching Revisited. Sig Proces Let, IEEE 2006; 13(5):285-287.

[13] Roy R, Sarkar A, Changder S. Chaos Based Edge Adaptive Image Steganography. Procedia Tech 2013;10(1):138-146.

[14] Hong W. Adaptive Image Data Hiding in Edges Using Patched Reference Table and Pair-Wise Embedding Technique. Info Sci 2013;221(1):473-489.

[15] Chen W-J, Chang C-C, Le T. High Payload Steganography Mechanism Using Hybrid Edge Detector. Exp Sys App 2010;37(4):3292-3301.

[16] Ioannidou A, Halkidis ST, Stephanides G. A Novel Technique for Image Steganography Based on a High Payload Method and Edge Detection. Exp Sys App 2012;39(14):11517-11524.

[17] Luo W, Huang F, Huang J. Edge Adaptive Image Steganography Based on LSB Matching Revisited. IEEE Trans Info Forens Sec 2010;5(2):201-214.

[18] Wang C-M, Wu N-I, Tsai C-S, Hwang M-S. A High Quality Steganographic Method with Pixel-value Differencing and Modulus Function. J Sys Soft 2008;81(1):150-158.

[19] Wu D-C, Tsai W-H. A Steganographic Method for Images by Pixel-value Differencing. Pat Recog Let 2003;24(9-10):1613-1626.

[20]  Gutub A, Ankeer M, Abu-Ghalioun M, Shaheen A, Alvi A. Pixel Indicator High Capacity Technique for RGB Image Based Steganography. In: 5th IEEE International Workshop on Signal Processing and its Applications. Sharjah, U.A.E: 2008. p. 1-3.

[21]  Gutub A A-A. Pixel Indicator Technique for RGB Image Steganography. J Emerg Tech Web Intel 2010;2(1):56-64.

[22]  Parvez MT, Gutub A-A. RGB Intensity Based Variable-Bits Image Steganography. In: IEEE Asia-Pacific Services Computing Conference (APSCC'08), Yilan, Taiwan: 2008. pp. 1322-1327.

[23]  Amirtharajan R, Behera SK, Swarup MA, Rayappan JBB. Colour Guided Colour Image Steganography. Universal Journal of Computer Science and Engineering Technology, 2010; 1(1): 16-23.

[24]  Parvez MT, Gutub A A-A. Vibrant Color Image Steganography Using Channel Differences and Secret Data Distribution. Kuwait J Sci Eng 2011;38(1B):127-142.

[25]  Amirtharajan R, Archana P, Rajesh V, Devipriya G, Rayappan J. Standard Deviation Converges for Random Image Steganography. In: IEEE Conference on Information & Communication Technologies (ICT). Tamil Nadu, India:, 2013. p. 1064-1069.

[26]  Amirtharajan R, Mahalakshmi V, Nandhini J, Kavitha R, Rayappan J. Key Decided Cover for Random Image Steganography. Res J Info Tech 2013;5(2):171-180.

[27]  Swain G, Lenka SK. A Novel Approach to RGB Channel Based Image Steganography Technique. Int. Arab J e-Tech 2012;2(4):181-186.

[28]  Mehmood I, Sajjad M, Ejaz W, Baik SW. Saliency-directed Prioritization of Visual Data in Wireless Surveillance Networks. Info Fusion 2015;24(1):16-30.

[29]  Sajjad M, Mehmood I, Abbas N, Baik SW. Basis Pursuit Denoising-based Image Superresolution Using a Redundant Set of Atoms. Sig, Imag Vid Process 2014;8(8):1-8.

[30] Ahmad J, Sajjad M, Mehmood I, Rho S, Baik SW. Describing Colors, Textures and Shapes for Content Based Image Retrieval-A Survey. Journal of Platform Technology 2014; 2(4): 34-48.

[31] Chen W-Y. Color Image Steganography Scheme Using DFT, SPIHT Codec, and Modified Differential Phase-Shift Keying Techniques. App Math Comp 2008;196(1):40-54.

[32] Ghasemi E, Shanbehzadeh J, Fassihi N. High Capacity Image Steganography Based on Genetic Algorithm and Wavelet Transform. In: Intelligent Control and Innovative Computing, Springer, 2012. p. 395-404.

[33] Raja K, Kumar K, Kumar S, Lakshmi M, Preeti H, Venugopal K. Genetic Algorithm Based Steganography Using Wavelets. In: Intelligent Control and Innovative Computing, Springer, 2007. p. 51-63.

[34] Qazanfari K, Safabakhsh R. High-capacity Method for Hiding Data in the Discrete Cosine Transform Domain. J Electro Imag 2013;22(4):043009(1)-043009 (10).

[35] Fakhredanesh M, Rahmati M, Safabakhsh R. Adaptive Image Steganography Using Contourlet Transform. J Electro Imag 2013;22(4):043007(1)-043007(13).

[36] Hussain M, Hussain M. A Survey of Image Steganography Techniques. Int J Advan Sci Tech 2013;54(2013):113-124.

[37] Li B, He J, Huang J, Shi YQ. A Survey on Image Steganography and Steganalysis. J Info Hiding Multi Sig Process 2011;2(2):142-172, 2011.

[38] Muhammad K, Sajjad M, Mehmood I, Rho S, Baik S. A Novel Magic LSB Substitution Method (M-LSB-SM) Using Multi-Level Encryption and Achromatic Component of an Image. Multi Tool App 2015; (10.1007/s11042-015-2671-9):1-27.

[39] Bailey K, Curran K. An Evaluation of Image based Steganography Methods. Multi Tool App 2006;30(1):55-88.

[40] Karim M. A New Approach for LSB Based Image Steganography Using Secret Key. In: 14$^t$h International Conference on Computer and Information Technology (ICCIT 2011). Dhaka, Bangladesh: 2011. p. 286-291.

[41] Fang Y, Zeng K, Wang Z, Lin W, Fang Z, Lin C-W. Objective Quality Assessment for Image Retargeting Based on Structural Similarity. IEEE J Emerg Sel Topics Circuit Syst 2014;4(1):95-105.

[42] Mehmood I, Sajjad M, Baik SW. Mobile-Cloud Assisted Video Summarization Framework for Efficient Management of Remote Sensing Data Generated by Wireless Capsule Sensors. Sens 2014;14(9):17112-17145.

[43] Sajjad M, Ejaz N, Mehmood I, Baik SW. Digital Image Super-resolution Using Adaptive Interpolation Based on Gaussian Function. Multi Tool App 2013; 74(20): 8961-8977.

[44] Muhammad JAK, Rehman NU, Jan Z. A Secure Cyclic Steganographic Technique for Color Images using Randomization. Tech J Uni Engg Tech Taxila 2014;19(3):57-64.

**Table 1. Comparison of methods using PSNR and MSE with different images**

| S. No | Image name | Karim et al. [40] method | Proposed method | Karim et al. [40] method | Proposed method | Karim et al. [40] method | Proposed method |
|---|---|---|---|---|---|---|---|
| | | PSNR (dB) | | MSE | | RMSE | |
| 1 | Baboon | 50.8811 | 58.0648 | 0.5121 | 0.4487 | 0.7156 | 0.6698 |
| 2 | Lena | 55.6551 | 58.0362 | 0.4682 | 0.4490 | 0.6842 | 0.6700 |
| 3 | Peppers | 17.3893 | 58.0362 | 1.4984 | 0.4490 | 1.2240 | 0.6700 |
| 4 | Building | 55.1595 | 59.3242 | 0.4724 | 0.4392 | 0.6873 | 0.6627 |
| 5 | Parrot | 41.9414 | 59.3242 | 0.6212 | 0.4392 | 0.7881 | 0.6627 |
| 6 | Army | 55.6788 | 60.3252 | 0.4680 | 0.4319 | 0.6841 | 0.6571 |
| 7 | Office | 45.3351 | 58.3514 | 0.5747 | 0.4465 | 0.7580 | 0.6682 |
| 8 | F16jet | 47.335 | 58.0576 | 0.5505 | 0.4488 | 0.7419 | 0.6699 |
| 9 | House | 50.949 | 58.1054 | 0.5114 | 0.4484 | 0.7151 | 0.6696 |
| 10 | Building1 | 28.7007 | 58.0116 | 0.9078 | 0.4491 | 0.9527 | 0.67014 |
| 11 | Trees1 | 38.7399 | 58.1752 | 0.6726 | 0.4479 | 0.8201 | 0.6692 |
| 12 | Trees2 | 24.0736 | 58.0284 | 1.0823 | 0.4490 | 1.0403 | 0.6700 |
| 13 | Girl1 | 28.3517 | 58.1558 | 0.9190 | 0.4480 | 0.9586 | 0.6693 |
| 14 | Girl2 | 55.9034 | 58.1872 | 0.4661 | 0.4478 | 0.6827 | 0.6691 |
| 15 | Girl3 | 50.1731 | 59.595 | 0.5193 | 0.4372 | 0.7206 | 0.6612 |
| | Average | 40.5017 | 58.5185 | 0.6829 | 0.4453 | 0.8115 | 0.6673 |

**Table 2. Comparison of methods using PSNR with variable amount of embedded cipher**

| Image name | Cipher size in (KBs) | cipher size in bytes | cipher size in bits | Karim et al. [40] method | Proposed method |
|---|---|---|---|---|---|
| | | | | PSNR (dB) | PSNR (dB) |
| baboon with dimension 256×256 | 2 | 2406 | 19248 | 52.0373 | 65.9333 |
| | 4 | 4177 | 33416 | 51.6345 | 60.8388 |
| | 6 | 6499 | 51992 | 51.1776 | 59.0243 |
| | 8 | 8192 | 65536 | 50.8811 | 58.0648 |

**Table 3. Comparison of both methods using PSNR with variable image dimensions**

| Image name | Cipher embedded (bits) | Image dimensions | Karim et al. [40] method | Proposed method |
|---|---|---|---|---|
| | | | PSNR (dB) | PSNR (dB) |
| Baboon | 1720 | 128×128 | 65.5328 | 67.9197 |
| | 1720 | 256×256 | 50.8811 | 74.1737 |
| | 1720 | 512×512 | 37.2456 | 88.0026 |
| | 1720 | 1024×1024 | 41.9577 | 85.8083 |

**Table 4. Comparison of both methods using NCC with different images**

| Serial No | Image Name | Karim et al. [40] method | Proposed Method |
|---|---|---|---|
| | | NCC | NCC |
| 1 | Baboon | 0.9998 | 0.9999 |
| 2 | Lena | 0.9999 | 0.9999 |
| 3 | Peppers | 0.7859 | 0.9093 |
| 4 | Building | 0.9999 | 0.9999 |
| 5 | Parrot | 0.9991 | 0.9995 |
| 6 | Army | 0.9999 | 0.9999 |
| 7 | Office | 0.9998 | 0.9999 |
| 8 | F16jet | 0.9997 | 0.9998 |
| 9 | House | 0.9998 | 0.9999 |
| 10 | Building1 | 0.9796 | 0.9899 |
| 11 | Trees1 | 0.9989 | 0.9994 |
| 12 | Trees2 | 0.9721 | 0.9897 |
| 13 | Girl1 | 0.9809 | 0.9898 |
| 14 | Girl2 | 0.9999 | 0.9999 |
| 15 | Girl3 | 0.9998 | 0.9999 |
| | Average | 0.981 | 0.9917 |

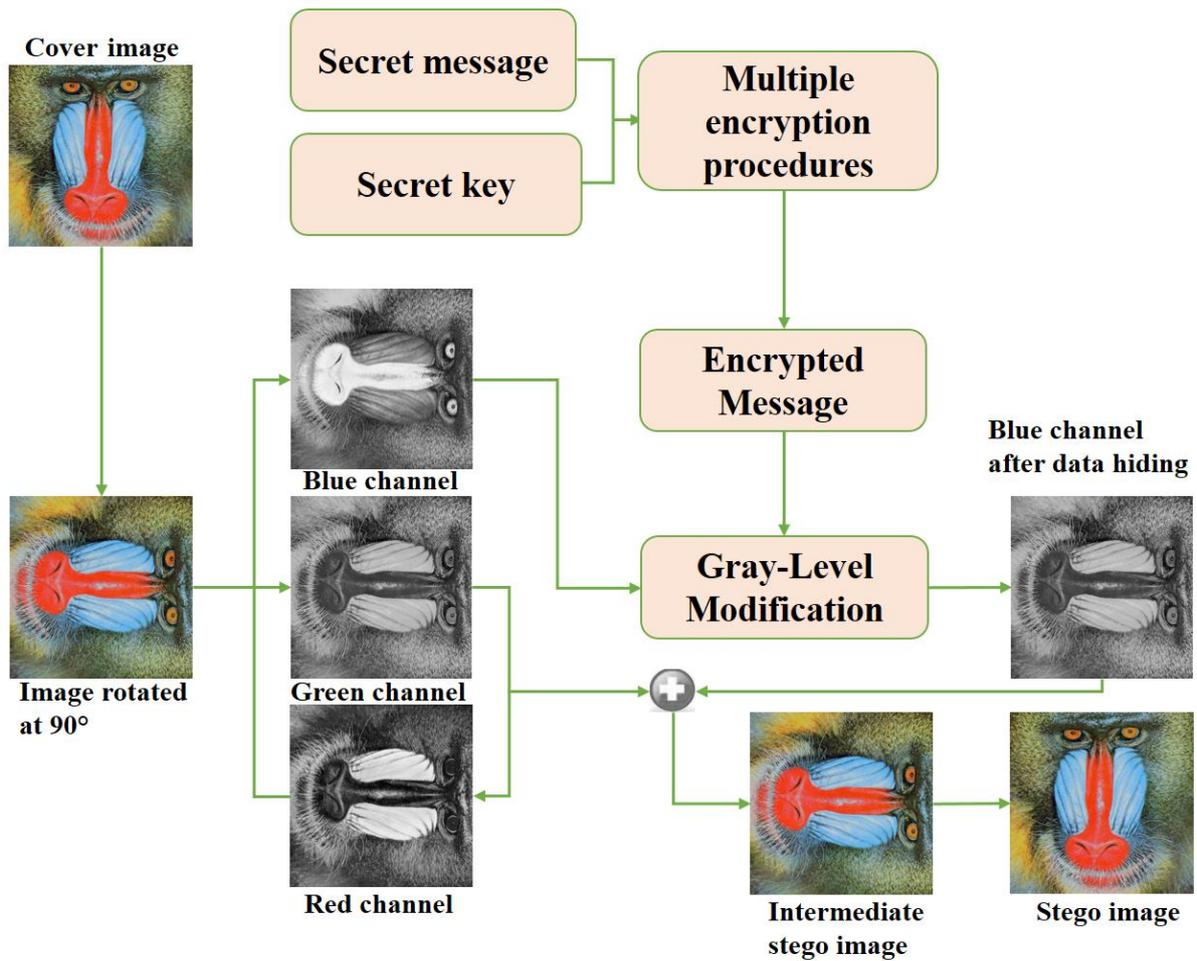

**Figure 1. Overall pictorial representation of proposed framework.**

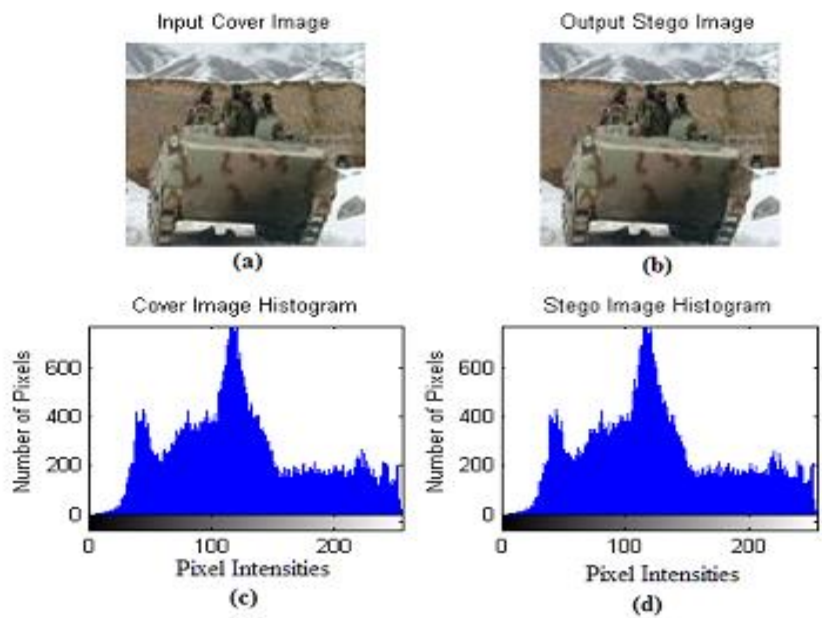

**Figure 2. Army tank input and output image with their histograms.**

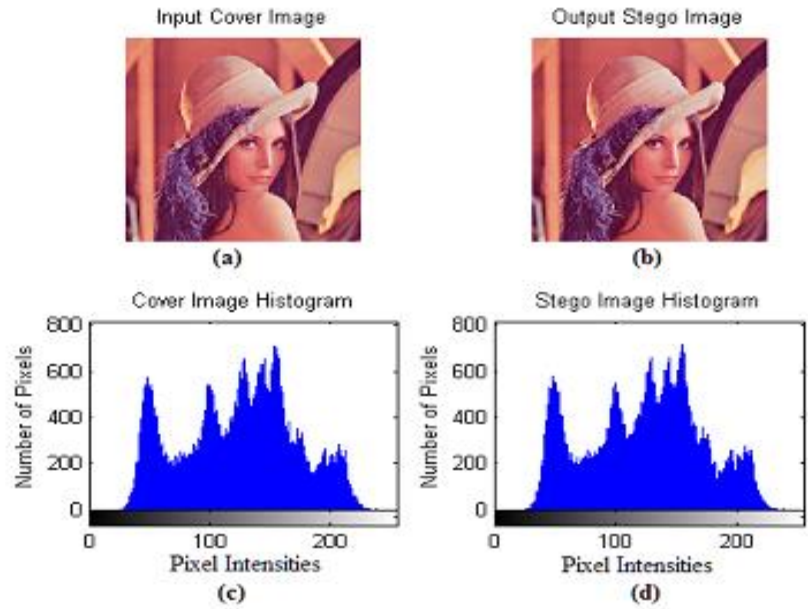

**Figure 3. Lena input and output image with their histograms.**

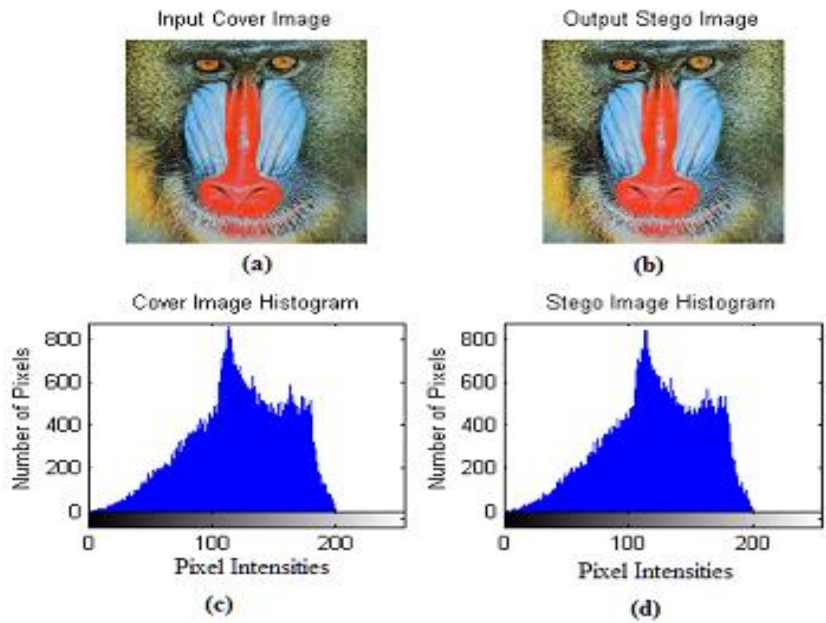

**Figure 4. Baboon input and output image with their histograms.**

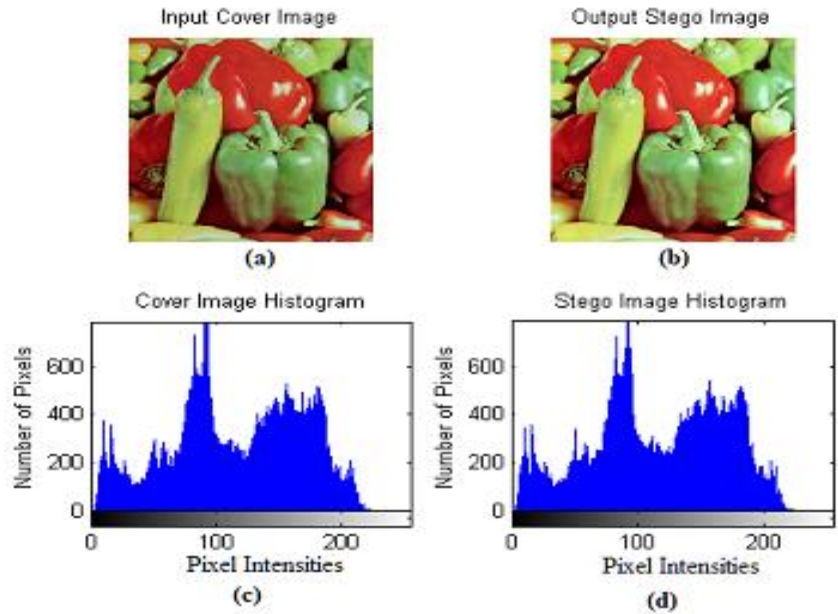

**Figure 5. Peppers input and output image with their histograms.**

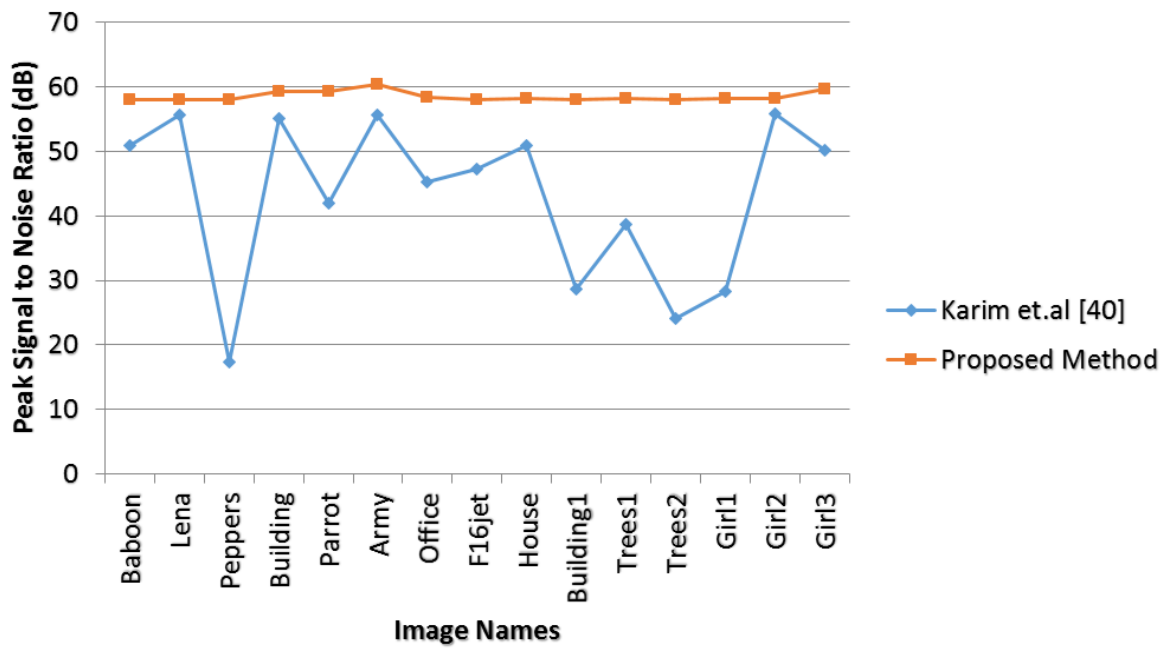

**Figure 6. Comparative Analysis of both methods using PSNR with different images.**

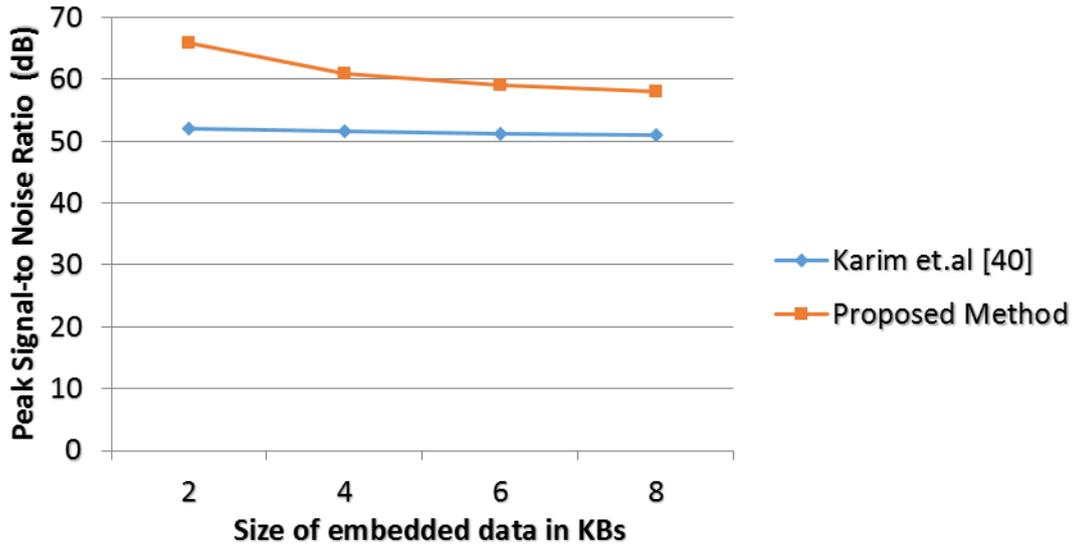

**Figure 7. Comparative analysis using PSNR with variable amount of embedded cipher.**

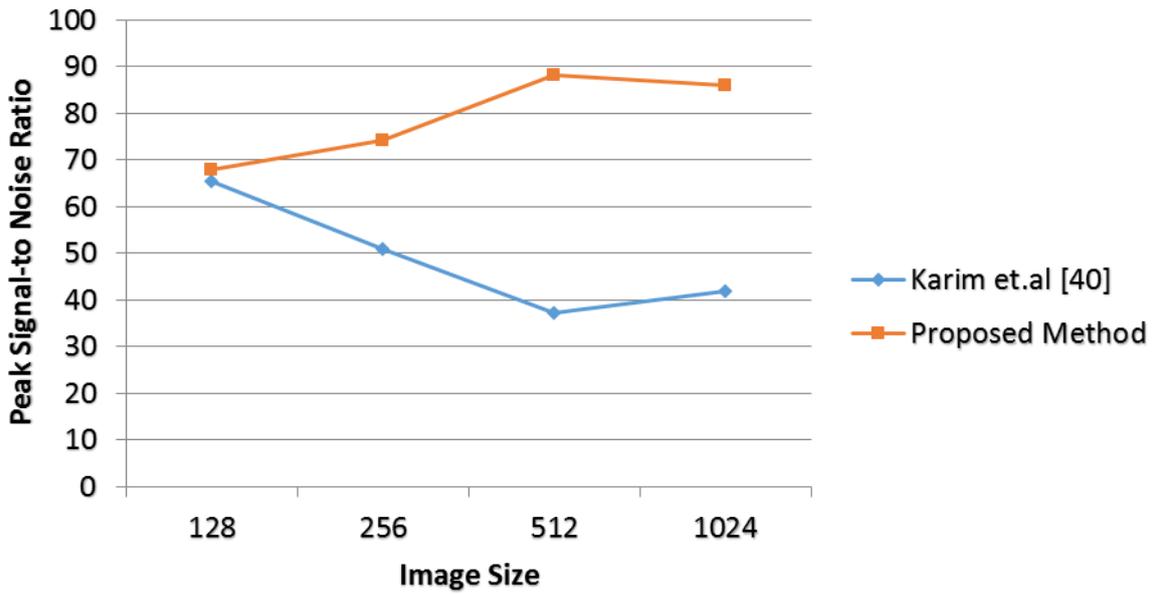

**Figure 8. Comparative analysis using PSNR with variable image dimensions.**